\documentclass[aps,pra,article,superscriptaddress,twocolumn]{revtex4} 
\usepackage{graphicx}
\usepackage{amsmath}
\usepackage{color}
\usepackage{marginnote}
\usepackage{tikz}

\begin{document}
\title{Quantum estimation via parametric amplification in circuit QED arrays}

\author{Ashley Wilkins}
\affiliation{School of Mathematical Sciences, University of Nottingham,
  NG7 2RD Nottingham, United Kingdom}
\email{ppyaw2@nottingham.ac.uk}
 \author{Carlos Sab\'in}
 \affiliation{School of Mathematical Sciences, University of Nottingham,
  NG7 2RD Nottingham, United Kingdom}
 \affiliation{Instituto de F\'isica Fundamental, CSIC, Serrano 113-bis
  28006 Madrid, Spain}

\email{csl@iff.csic.es}

  \date{\today}

\begin{abstract}
We propose a scheme for quantum estimation by means of parametric amplification in circuit Quantum Electrodynamics. The modulation of a SQUID interrupting a superconducting waveguide transforms an initial thermal two-mode squeezed state in such a way that the new state is sensitive to the features of the parametric amplifier. We find the optimal initial parameters which maximize the Quantum Fisher Information. In order to achieve a large number of independent measurements we propose to use an array of non-interacting resonators. We show that the combination of both large QFI and large number of measurements enables, in principle, the use of this setup for Quantum Metrology applications. 
\end{abstract}
\maketitle
\section{Introduction}
Circuit Quantum Electrodynamics (QED) \cite{reviewdevoret,reviewnori} has quickly developed in the last decade and is now one of the most promising platforms for quantum technologies such as quantum computers \cite{reviewwilhelm,annealing} and quantum simulators \cite{cavity,simulatorfermion} due to, among other reasons, the high level of controlability and scalability that can be achieved. In circuit QED both superconducting qubits and electromagnetic radiation can be controlled and manipulated to an extent that goes beyond some of the standard restrictions in other platforms of quantum optics and quantum information.

An important area in quantum technologies is devoted to the emergent field of quantum metrology \cite{advances} which aims at improving the precision of measurement devices by exploiting quantum features such as entanglement and squeezing in phase estimation protocols. Applications are critical and diverse, ranging from the use of squeezed states in gravitational wave detection with laser interferometers \cite{ligogw} to the notion of a global network of quantum clocks \cite{komarnet}, among many others.

In this work we propose a scheme for quantum  estimation in circuit QED. We consider a superconducting transmission line interrupted by a superconducting quantum interference device (SQUID). This technology resembles the one employed in the observation of the Dynamical Casimir Effect \cite{casimirwilson}. However, instead of an initial vacuum we consider the preparation of more general initial states- in particular thermal two-mode squeezed states- by means of an additional transmission line. The modulation of the SQUID transforms this initial state in such a way that it becomes dependent on the parameters of the modulating magnetic field. Thus the parameters of the magnetic field can be estimated by means of  phase estimation techniques. We compute the Quantum Fisher Information (QFI) and maximize it over the set of considered states in order to determine the optimal initial estate for quantum estimation. In order to maximise as well the number of independent measurements and accordingly the precision, instead of considering a single superconducting resonator we propose the use of a large array of non-interacting cavities \cite{cavity}. We show that good precision can be achieved for realistic experimental parameters.  We discuss possible applications of these results, which include accurate frequency measurements or highly precise measurements of magnetic flux variations threading the SQUID.

The structure of the paper is the following. In Section 2 we introduce our model and show how to maximize the QFI of the electromagnetic field state confined within a single superconducting resonator. In section 3 we show how these results enable the use of an array of resonators for quantum estimation, discussing some potential applications. We conclude in Section 4 with a summary of our results

\section{Quantum Fisher Information in a single superconducting waveguide}

We will consider a large array of superconducting resonators \cite{cavity} consisting of superconducting waveguides terminated by SQUIDs \cite{casimirtheory,casimirwilson} with mutual interactions controlled by additional SQUIDs \cite{liberatoarray}. As we will see in detail below we want to achieve a large number of independent measurements of a particular quantum state of the electromagnetic field. To this end, the SQUIDs can be tuned in a such a way that the resonators are non-interacting \cite{borjatunable}. Therefore, we can focus in the dynamics of a single superconducting waveguide as we will do in the following. 

In order to exploit squeezing and entanglement we consider the preparation of  a two-mode squeezed state as initial state. This can be achieved by connecting the transmission line to an auxiliary line terminated by an array of three SQUIDs which provide a Kerr medium that can be used as a parametric amplifier -this has been used to experimentally generate two-mode squeezed states within a single transmission line \cite{wallraffsqueezing}. We consider as well a non-zero small temperature characterized by a small number of thermal photons $n^{th}$.

We will describe the field dynamics by means of the covariance matrix $V$. Using the same convention as in \cite{nonclassicaldce}, which assumes zero displacement without any loss of generality, we have $V_{\alpha\beta} = \frac{1}{2}\left<R_\alpha R_\beta+R_\beta R_\alpha\right>,$  where $R^{\rm T} = \left(q_-, p_-, q_+, p_+\right)$ is a vector with the quadratures as elements: $q_\pm = (b_\pm + b_\pm^\dag)/\sqrt{2}$ and $p_\pm = -i(b_\pm - b_\pm^\dag)/\sqrt{2},$ given in terms of the creation $b_\pm^\dag$ and annihilation $b_\pm$ operators of the two modes of interest $+,-$. 

The initial state is then described by the covariance matrix of a thermal two-mode squeezed state:
\begin{eqnarray}
V &=&\dfrac{1}{2} \begin{pmatrix} A & B\\ B & A\end{pmatrix} \nonumber\\
A &=&\operatorname{cosh}(2r)(1+2\,n^{\operatorname{th}}) \openone,\nonumber\\ 
B &=&\operatorname{sinh}(2r)(1+2\,n^{\operatorname{th}})\cos(\theta)\sigma_z \nonumber\\
&+& \operatorname{sinh}(2r)(1+2\,n^{\operatorname{th}})\sin(\theta)\sigma_x\nonumber
\end{eqnarray}
where $r$ and $\theta$ define the complex squeezing parameter $\chi=r\,e^{i\theta}$ and $\sigma_x,\sigma_z$ are standard Pauli matrices.\\

The aim now is to transform this initial state under the parametric amplification process induced by the modulation of the SQUID that terminates the waveguide. If we add a weak harmonic drive to the SQUID characterized by a frequency $\omega_d$ and a normalised amplitude $\epsilon$, then the field quadratures are transformed as follows \cite{nonclassicaldce}:
$q_\pm=-(q_{0\pm}+\,f\,p_{0\mp})$,
$p_\pm=-(p_{0\pm}+\,f\,q_{0\mp})$
where the small parameter $f$ is:
\begin{equation}
\label{eq:f}
f = \dfrac{\epsilon L_{\operatorname{eff}}\omega _{d}}{2v}.
\end{equation}
 $L_{\operatorname{eff}}$ is an effective length that describes the boundary conditions that the SQUID provides to the flux field while v is the speed of light along the waveguide. We are assuming that the frequencies of the modes are very close $\omega_+\simeq\omega_-\simeq\omega_d/2$.

Under this transformation and  considering only up to linear terms in $n^{\operatorname{th}}$ and up to quadratic terms in $f$ we obtain the transformed covariance matrix \~{V} of the state:
\begin{eqnarray}\label{eq:state}
\text{\~{A}} &=& \operatorname{cosh}(2r)(1+2n^{\operatorname{th}})(1+f^2 +2f\tanh(2r)\sin(\theta)) \openone\nonumber\\ 
\text{\~{B}}&=& \sinh(2r)(1+2n^{\operatorname{th}})(1-f^2)\cos(\theta)\sigma_z\nonumber\\
&+&[2f\cosh(2r)(1+2n^{\operatorname{th}}) + \nonumber\\
 & & (1 + f^2 + 2n^{\operatorname{th}})\sinh(2r)\sin(\theta)]\sigma_x.
\end{eqnarray}

Our main aim is to analyze the sensitivity of the state in Eq. (\ref{eq:state}) with respect to the parameter $f$.  In order to achieve this goal we consider the QFI, which provides a bound on the error of the estimation of the parameter. Therefore, we will seek to maximize the QFI. 

\subsection{Single-mode reduced covariance matrix}
First, let us analyze the case in which we try to estimate the parameter by means of measurements over only one mode. Therefore, we consider the reduced single-mode covariance matrix, that is $\text{\~{A}}$. 

The QFI for estimation of a parameter $\tau$ using a single-mode Gaussian state $\sigma$ is given in \cite{braun} and for zero displacement reduces to:
\begin{eqnarray}
H_{\tau} = \dfrac{1}{2}\dfrac{\operatorname{Tr}[(\sigma^{-1}(\tau)\sigma'(\tau))^{2}]}{1 + P(\tau)^{2}} + 2\dfrac{P'(\tau)^{2}}{1 - P(\tau)^{4}}, 
\end{eqnarray}
where P = $1/(4\sqrt{\operatorname{Det}\sigma})$ is the purity of the state and $\operatorname{Det}$ stands for the determinant of a matrix. 
In our case $\sigma$ is the reduced matrix \~{A}. The prime indicates a derivative with respect to the parameter $\tau$ (e.g. $P'(\tau) = \partial_\tau P$). For our purposes this parameter will be $f$. 
\begin{figure}
\includegraphics[scale=0.6]{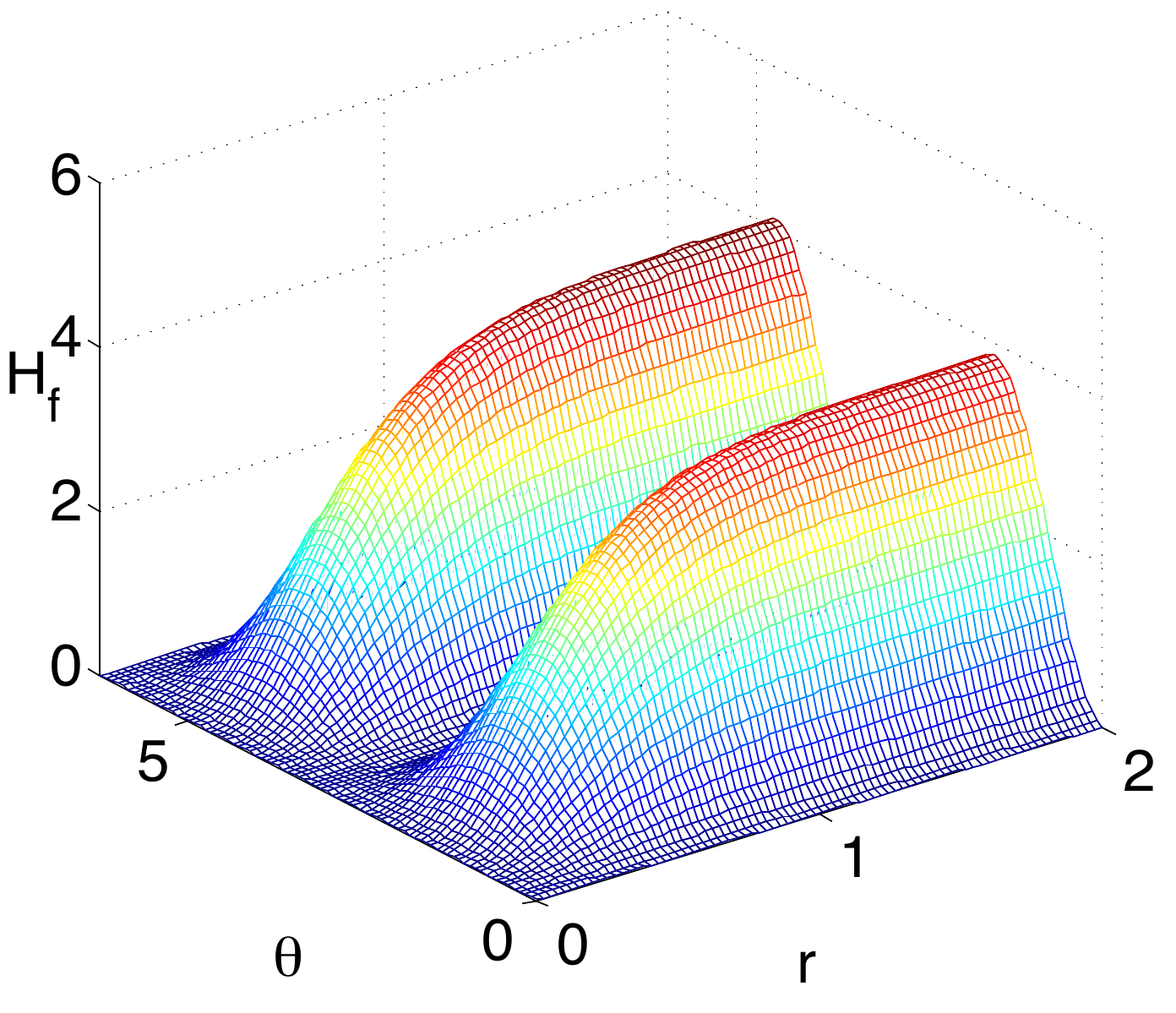}
\caption{(Color online) Single-mode QFI $(H_f)$ vs. $r$ and $\theta$ for $n^{th} = 8\cdot10^{-3}$ and $f \approx 0.02$. The QFI is maximized at $\theta= \pi/2$ and $\theta=3\pi/2$ and grows monotonically with $r$.}
\label{fig:0}
\end{figure}

In Fig. (\ref{fig:0}) we plot $H_f$ with respect to $r$ and $\theta$, using realistic experimental parameters $n^{\operatorname{th}}=8\cdot10^{-3}$-which corresponds to a temperature $T=50\operatorname{mK}$-, $\omega_d=2\pi \times 10 \operatorname{GHz}$, $L_{\operatorname{eff}}=0.4 \operatorname{mm}$ and $\epsilon=0.25$ \cite{casimirwilson, casimirtheory}. We find that the QFI oscillates with $\theta$ in such a way that the maximum is reached at $\theta=\pi/2, 3\pi/2$ while the minimum is at $\theta=0, \pi$. The QFI grows significantly with the value of the squeezing parameter $r$ -as expected- which in the figure is plotted in a realistic range $r<2$ \cite{wallraffsqueezing}.

\begin{figure}
\includegraphics[scale=0.6]{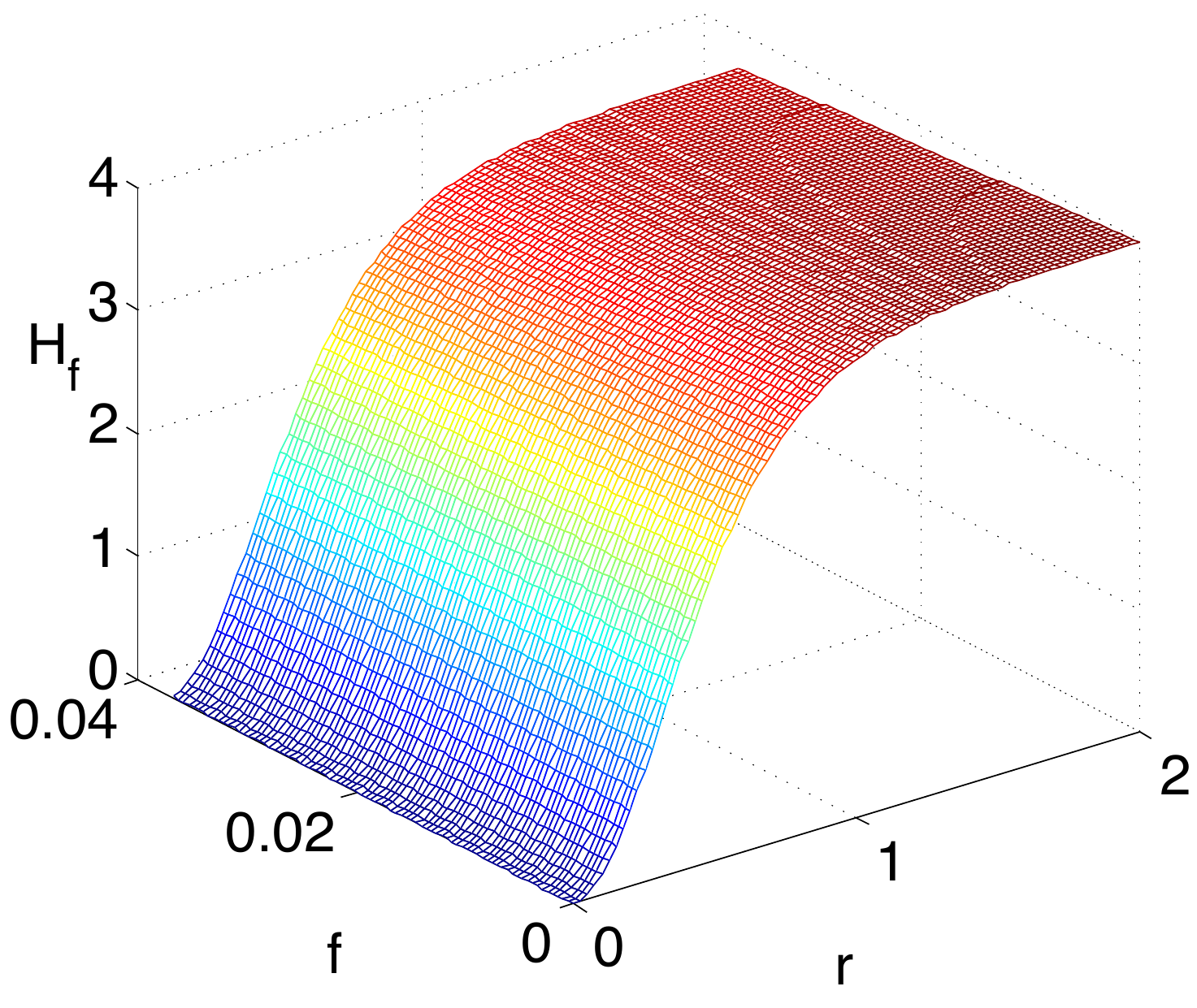}
\caption{(Color online) Single-mode QFI $(H_f)$ vs $r$ and $f$ for $n^{\operatorname{th}} = 8\cdot10^{-3}$ and the optimal value $\theta = \pi/2$. The QFI slightly diminishes with $f$ and grows dramatically with $r$. }
\label{fig:1}
\end{figure}

In Fig. (\ref{fig:1}) we choose the optimal value $\theta=\pi/2$ and plot the single-mode QFI vs. $r$ and $f$. We see that the QFI slightly decreases with $f$, while the growth of the QFI with $r$ is observed at any value of $f$.
We have also found that the QFI is highly insensitive to the value of $n^{\operatorname{th}}$ within the perturbative regime that we are considering here.

\subsection{Full two-mode state}

Now we analyze the case in which the full two-mode covariance matrix is used for the estimation protocol. The two-mode QFI $H_{\tau}$ with respect to a parameter $\tau$ can be computed by means of the Uhlmann fidelity in the following way \cite{adesso}: 
\begin{equation}
H_{\tau} = -2\lim_{\delta\to 0} \dfrac{\partial^2 F}{\partial^2 \delta} 
\end{equation} 
where $F$ is the Uhlmann fidelity given by \cite{marianmarian}:
$F = 1/\big( \sqrt{\Gamma} + \sqrt{\Lambda} -\sqrt{(\sqrt{\Gamma} + \sqrt{\Lambda})^2 - \Upsilon}\, \big) $
and where
$\Gamma = 16\operatorname{Det}[\Omega(\text{\~{V}}_{1}/2)\Omega(\text{\~{V}}_2/2)-\dfrac{1}{4}],
\Lambda = 16\operatorname{Det}[(\text{\~{V}}_1 + i\Omega)/2]\operatorname{Det}[(\text{\~{V}}_2 + i\Omega)/2],
\Upsilon = \operatorname{Det}[(\text{\~{V}}_1 + \text{\~{V}}_2)/2]$
$\Omega$ being the symplectic form $\Omega=i\,\sigma_x\oplus i\,\sigma_x$,
and the covariance matrices \~{V}$_{1}$ and \~{V}$_{2}$ only differ in an infinitesimal variation of the parameter of interest, that is \~{V}$_{1}$ depends on $\tau$ while \~{V}$_{2}$ depends on $\tau +\delta$. Thus, in our case, \~{V}$_{1}$ is given by Eq.(\ref{eq:state}) and \~{V}$_{2}$ is obtained by replacing $f$ by $f+\delta$.

Putting all together we are able to find a simple analytical expression for the leading order in perturbation theory of the two-mode QFI:
\begin{eqnarray}
\label{eq:QFI2}
H_f &=& 4[\sinh^2(2r)\cos^2(\theta)(1+4f^2-4n^{\operatorname{th}})\nonumber\\&-&f^2+n^{\operatorname{th}}(\sqrt{\dfrac{17}{2}} - 2)]\nonumber\\
\end{eqnarray}

\begin{figure}
\includegraphics[scale=0.6]{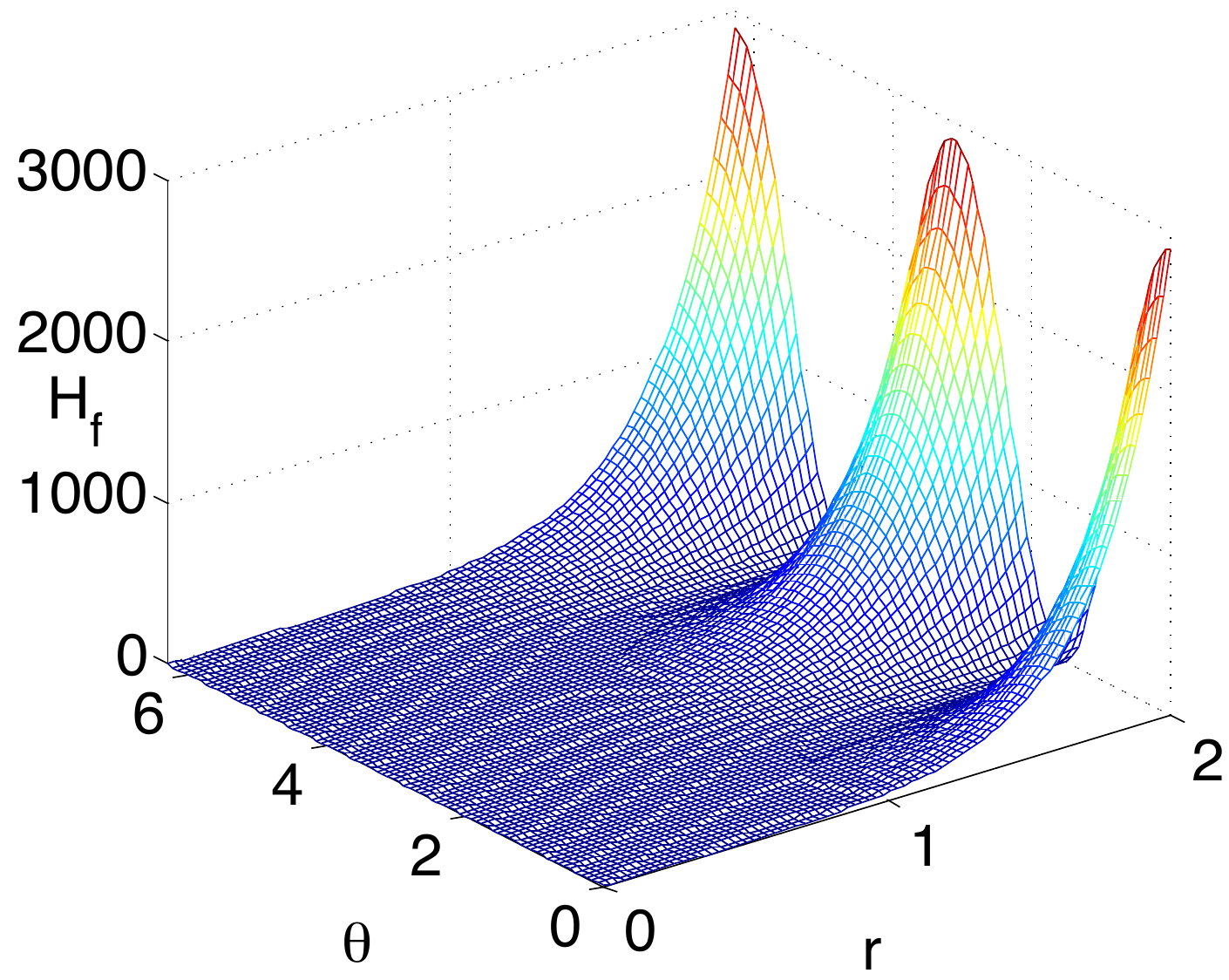}
\caption{(Color online) Two-mode QFI $(H_f)$ vs. $r$ and $\theta$ for $n^{\operatorname{th}} = 8\cdot10^{-3}$ and $f \approx 0.02$. The optimal values are $\theta=0$ and $\pi$ while the QFI grows monotonically with $r$. The QFI is three orders of magnitude larger than the single-mode case.}
\label{fig:2}
\end{figure}

\begin{figure}
\includegraphics[scale=0.4]{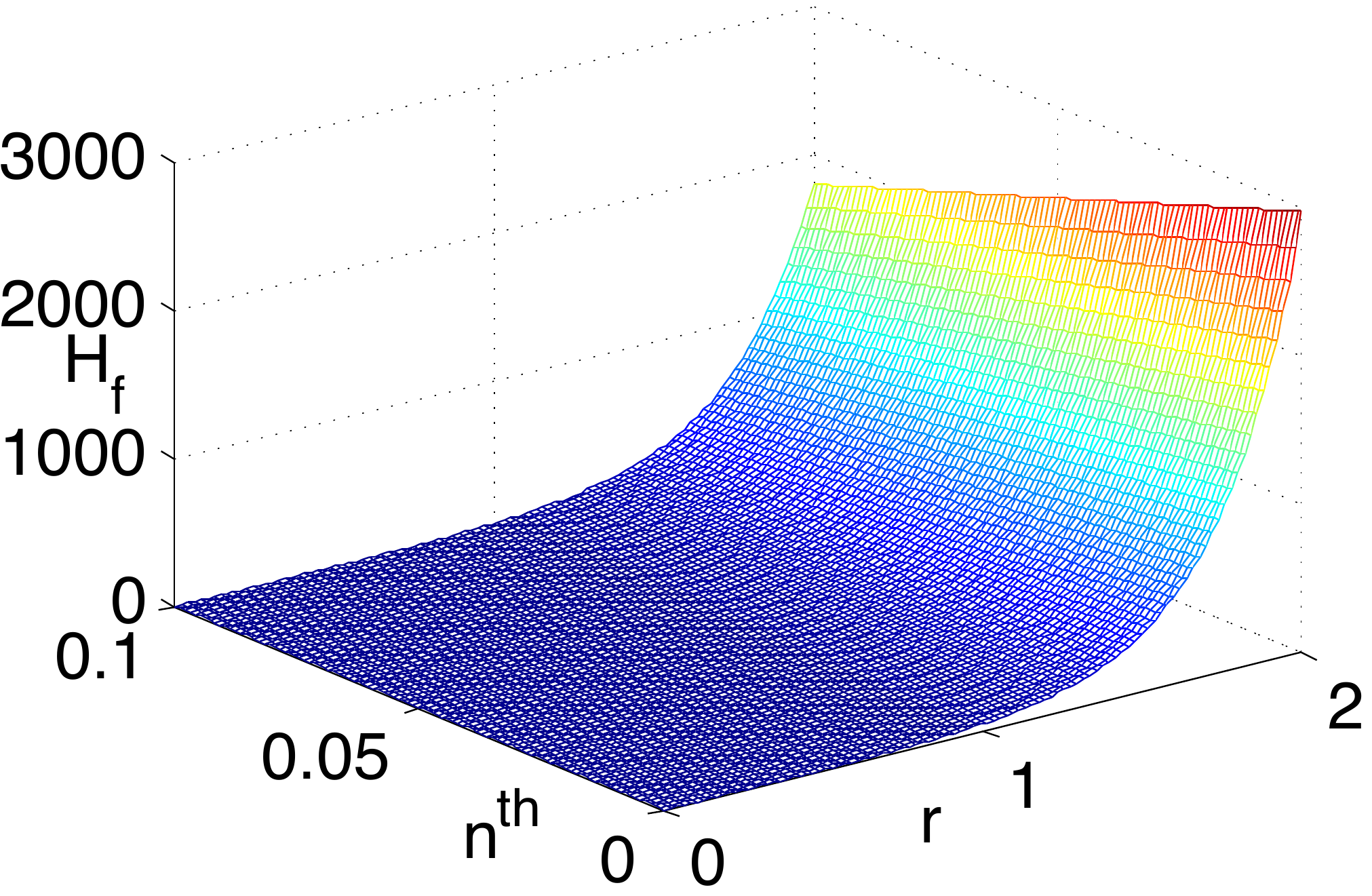}
\caption{(Color online) Two-mode QFI $(H_f)$ vs. $r$ and $n^{\operatorname{th}}$ for the optimal value $\theta = 0$ and $f \approx 0.02$. We see that thermal noise slightly degrades the QFI.}
\label{fig:n}
\end{figure}

\begin{figure}
\includegraphics[scale=0.6]{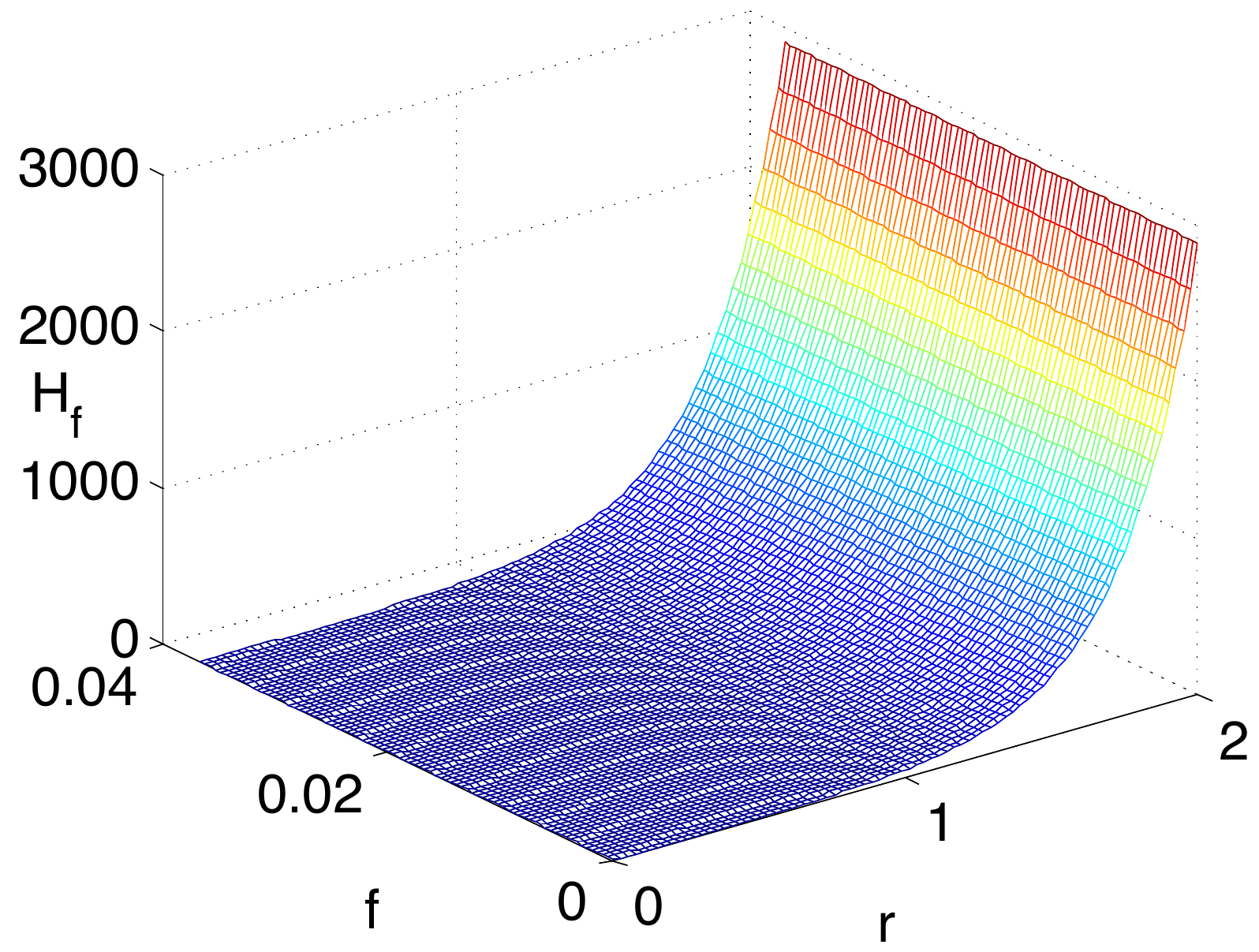}
\caption{(Color online) Two-mode QFI $(H_f)$ vs. $r$ and $f$ for the optimal value $\theta = 0$ and the experimental one $n^{\operatorname{th}} = 8\cdot10^{-3}$ -corresponding to $T=50 \operatorname{mK}$.The QFI is highly insensitive to the value of $f$ while grows significantly with $r$.}
\label{fig:3}
\end{figure}

In Fig.\ref{fig:2} we plot this two-mode QFI vs r and $\theta$ for the same parameters as in \ref{fig:0}. We see that the QFI for the optimal parameters is three orders of magnitude larger than the best scenario in the single-mode case. The QFI oscillates with $\theta$, but in this case $\theta=0, \pi$ are the optimal values. As expected, the QFI grows drastically with $r$. In Figs. \ref{fig:n} and \ref{fig:3} we see that the QFI is almost independent of the value of $f$ and slightly decreases with $n^{\operatorname{th}}$ -in both cases for the optimal value of $\theta$. 

In summary, the strategy to maximize the QFI would be to consider the joint two-mode state, to choose an optimal value of $\theta=0,\pi$ and to achieve a squeezing parameter $r$ as large as possible given the experimental limitations. 

In the next section, we will see how all the above is related with the error in the measurement of physical magnitudes.

\section{Quantum estimation of physical parameters in circuit QED arrays}

The relation between the optimal uncertainty in the estimation of $f$ and the QFI is governed by the quantum Cramer-Rao bound \cite{advances}:
\begin{eqnarray}\label{eq:cramerrao}
\Delta f\geq \dfrac{1}{\sqrt{M}\sqrt{H_f}} \nonumber\\\nonumber
\end{eqnarray}
where $M$ is  the number of independent measurements performed on the state. There always exists an optimal measurement strategy that saturates the bound.

In the previous section, we have analyzed how to maximize the QFI. In the following we discuss how to maximize the number of measurements. 

To this end we consider a large array of superconducting resonators \cite{cavity,cavity2} in which additional SQUIDs control the interaction between each resonator \cite{liberatoarray}.  In particular, these SQUIDs can be tuned in order to switch off the coupling \cite{borjatunable} and to obtain a lattice of non-interacting resonators. Thus we can assume that we have a large number of copies of the same individual superconducting resonator. In this way, we can use a large number $M$ in Eq. (\ref{eq:cramerrao}) in order to mimimize the error. In particular, a number of $10^3$ resonators seems to be within reach of current technology \cite{cavity}. 

\begin{figure}[h!]
\begin{center}
\includegraphics[width=0.45\textwidth]{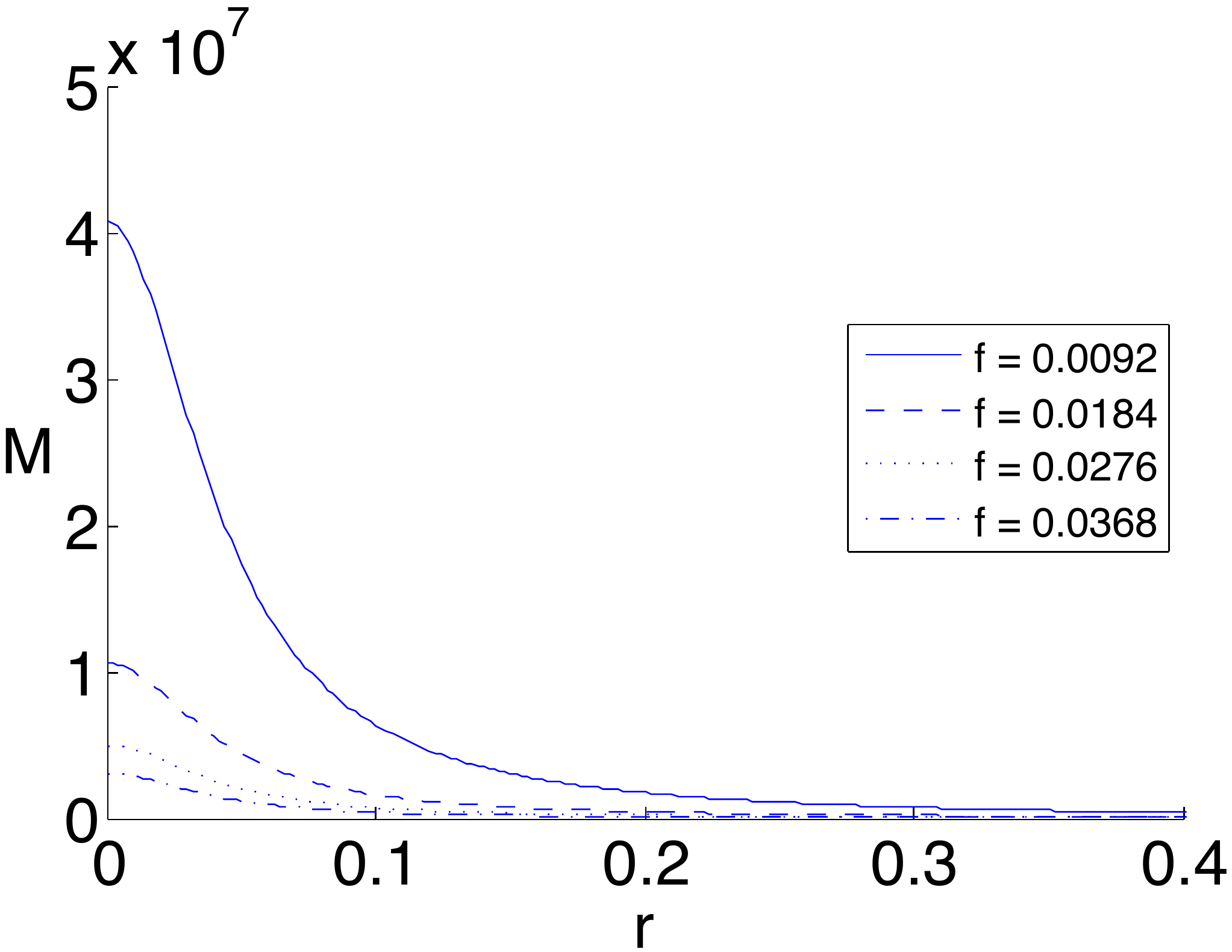}\\
\includegraphics[width=0.45\textwidth]{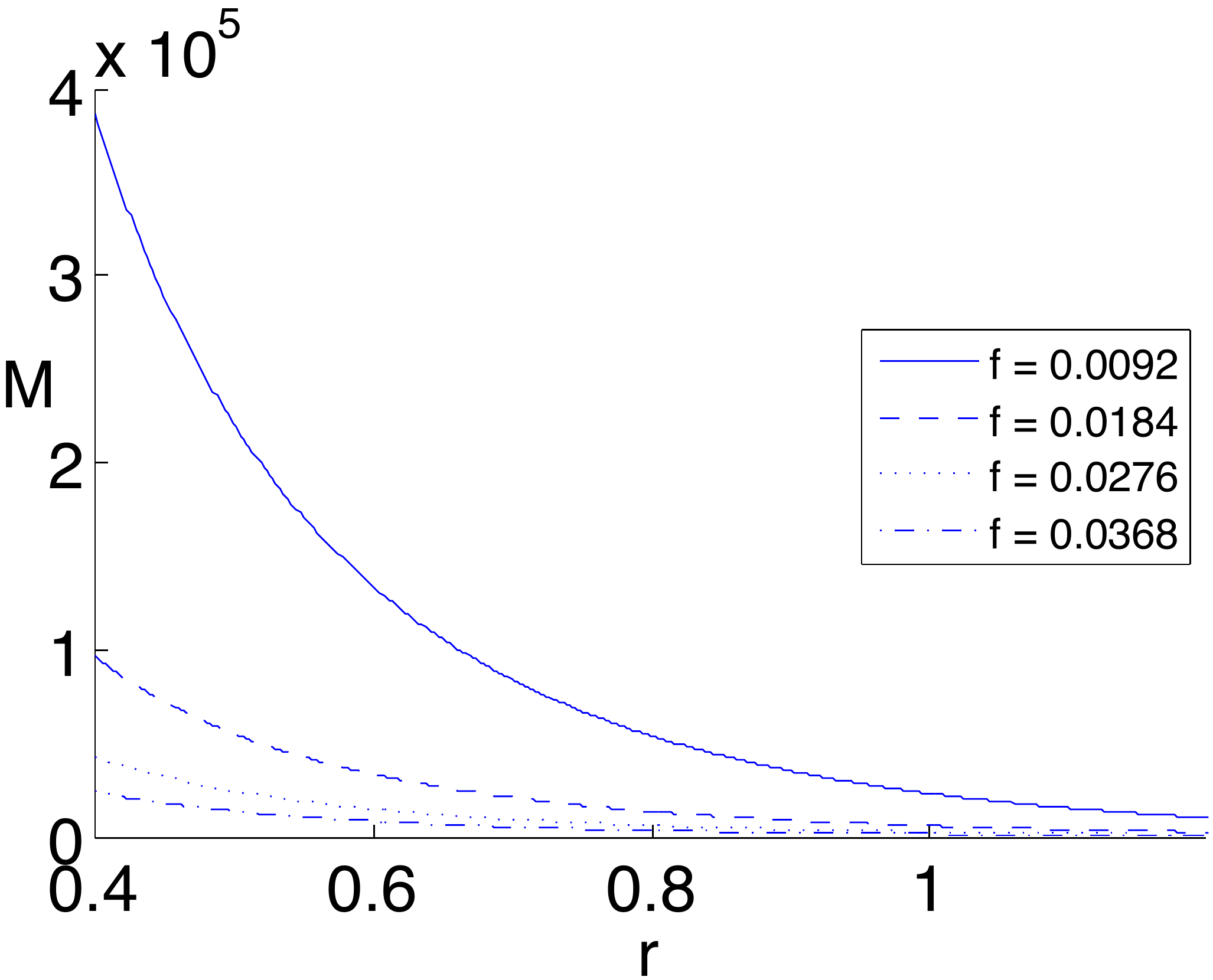}\\
\includegraphics[width=0.45\textwidth]{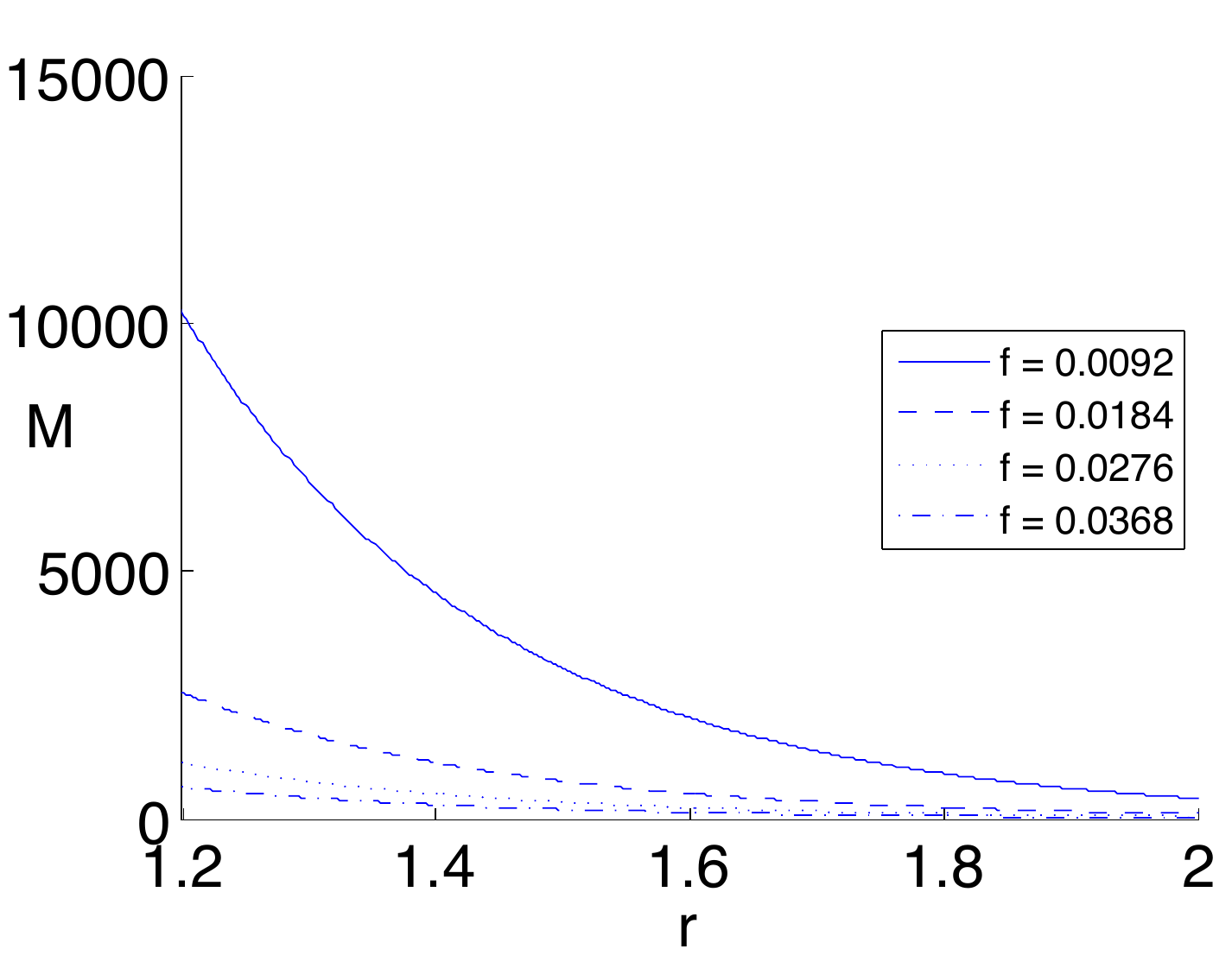}
\caption{(Color online) Number of independent measurements $M$ needed to achieve a relative error $E\leq 0.1$ in the estimation of $f$ vs. $r$, for different values of $f$, $n^{\operatorname{th}} = 8\cdot10^{-3}$ and $\theta = 0$.}
\label{figuremeasure}
\end{center}
\end{figure}

\begin{figure}
\includegraphics[scale=0.4]{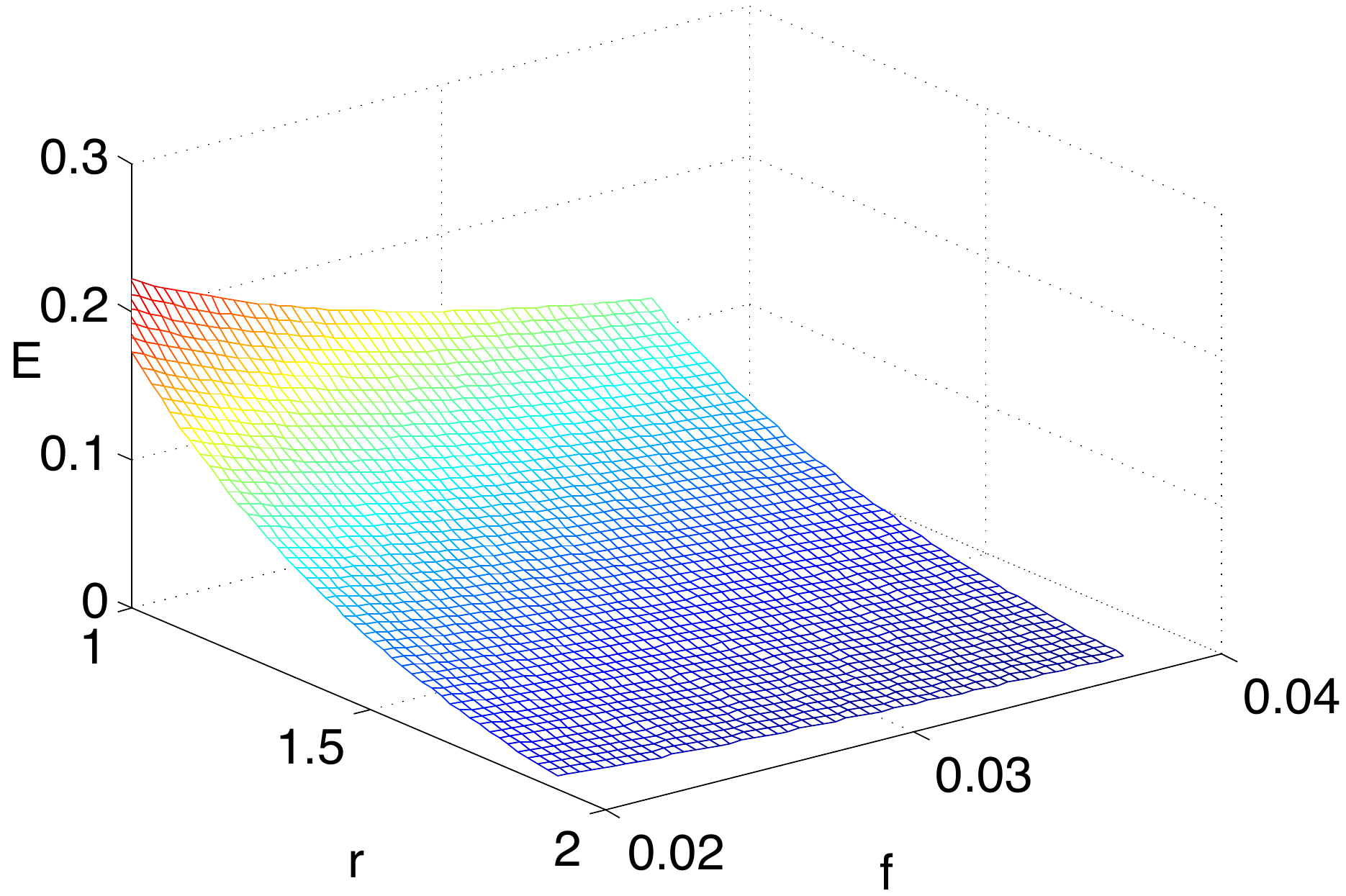}
\caption{(Color online) Error (E = $\Delta f/f$) for different values of $r$ and $f$ assuming $M=10^3$, $n^{\operatorname{th}} = 8\cdot10^{-3}$ and $\theta = 0$.}
\label{fig:err}
\end{figure}

In Fig. \ref{figuremeasure} we show the number of measurements required to achieve a relative error $E=\Delta f/f\leq 0.1$ for different values of the parameters $r$ and $f$. We see that for large enough values of $r$,  $M$ can be comparable to the desired reference value $10^3$, although the number grows for the lowest values of $f$. Indeed, in Fig. \ref{fig:err} we plot $E$ vs. $r$ and $f$ assuming $M=10^3$, showing that $E$ can be extremely small for the largest values of $r$.

The above results show that we can estimate with high precision the value of $f$, that is the degree of additional squeezing introduced by the parametric amplification. Perhaps more interestingly from the physical viewpoint, notice that $f$ is a product of several physical magnitudes $\omega_d$, $v$ and $L_{\operatorname{eff}}$ - Eq. (\ref{eq:f}). 
Thus we can relate the uncertainty in the estimation of f with their respective uncertainties via the standard error propagation formula:
\begin{eqnarray}\label{eq:relativeerr}
\dfrac{\Delta f}{f}&=& \sqrt{(\dfrac{\partial f}{\partial v}\dfrac{\Delta v}{f})^2 + (\dfrac{\partial f}{\partial \omega _{d}}\dfrac{\Delta \omega _{d}}{f})^2 + (\dfrac{\partial f}{\partial L}\dfrac{\Delta L}{f})^2}\nonumber\\
&=& \sqrt{(\dfrac{\Delta v}{v})^2 + (\dfrac{\Delta \omega _{d}}{\omega _{d}})^2 + (\dfrac{\Delta L}{L})^2},
\end{eqnarray}
where in the last line we have already used Eq. (\ref{eq:state}). Assuming that we can consider the scenario in which we have good control over all the variables but one -for instance, by means of a careful calibration process so that we can account for the uncertainties in the other variables as systematic errors- then we have that: $E=\Delta f/f=\Delta x/x$ -where we are denoting by $x$ the mentioned magnitude of interest. That is, the relative error in the estimation of $f$ is equal to the relative error in the estimation of $x$. Since we have already shown that we are able to achieve very high precision in the estimation of $f$, this entails that the same conclusion can be extended to the estimation of $\omega_d$, $v$ and $L_{\operatorname{eff}}$ -provided that it is possible to realise an experimental scenario in which we have control over all the variables but one. 

Moreover, $L_{\operatorname{eff}}$ can be related to the magnetic flux threading the SQUID. Indeed, the effective length is given by 
\begin{equation}\label{eq:effec}
L_{\operatorname{eff}}=(\dfrac{\phi_0}{2\pi})^2\dfrac{1}{E_J L_0}
\end{equation}
where $\phi_0$ is the magnetic flux quantum, $L_0$ the inductance per unit length of the superconducting waveguide and $E_J$ the flux-dependent Josephson energy of the SQUID, given by:
\begin{equation}\label{eq:joseph}
E_J=2 I_c \dfrac{\phi_0}{2\pi}|\cos(\dfrac{\phi_{\operatorname{ext}}}{\phi_0})|,
\end{equation}
where $I_c$ is the SQUID critical current and $\phi_{\operatorname{ext}}$ the external magnetic flux. By combining Eqs. (\ref{eq:effec}) and (\ref{eq:joseph}) and using an error propagation formula similar to Eq. (\ref{eq:relativeerr}) we find:
\begin{equation}\label{eq:fluxometer}
\dfrac{\Delta L_{\operatorname{eff}}}{L_{\operatorname{eff}}}=\pi\tan{\big(\dfrac{\pi\phi_{\operatorname{ext}}}{\phi_0}\big)}\dfrac{\delta \phi_{\operatorname{ext}}}{\phi_0}.
\end{equation}
In the experiments, $\phi_{\operatorname{ext}}\simeq 0.35\phi_0$ and thus we find that $\Delta L_{\operatorname{eff}}/L_{\operatorname{eff}}\simeq2\Delta \phi_{\operatorname{ext}}/\phi_{\operatorname{ext}}$. Therefore, a precise estimation of $f$ entails a precise estimation of the magnetic flux. Of course, highly accurate and sensitive magnetometers are already available and indeed SQUIDs are well-known as ultrasensitive magnetometers. A thorough investigation on whether our quantum metrology techniques can actually improve on the current state of the art lies beyond the scope of this work.

\section{Conclusions}

We have shown that quantum metrology tools can be used to estimate physical variables in a circuit QED scenario. In particular, we have considered a superconducting waveguide interrupted by a SQUID, where modulation of the magnetic field which threads the SQUID acts as a parametric amplifier. 

We start from a thermal two-mode squeezed state where the squeezing is characterized by the parameters $r$ and $\theta$. We find the optimal parameters that maximize the QFI and therefore the initial state which is more suitable for quantum phase estimation. After computing both the QFI of the full state and the reduced single-mode state, we conclude that the best strategy is to consider the full two-mode state where the initial parameters are $\theta=0,\pi$ and $r$ as large as allowed by the experimental limitations. In order to achieve a large number of independent measurements, we propose to use a large array of superconducting waveguides, where additional SQUIDs control the interaction strength in order to ensure a large number of non-interacting superconducting resonators, providing copies of the single-resonator system. We show that the combination of large QFI and large independent measurements enables a precise estimation of the parameter $f$, which characterizes the process of parametric amplification. This can be used for a precise estimation of the physical magnitudes involved in the definition of $f$, for instance the magnetic flux threading the SQUID.
\section*{Acknowledgements}
Financial support by Fundaci{\' o}n General CSIC (Programa ComFuturo) is acknowledged by CS. AW acknowledges funding of the Research Bursary program of the School of Mathematical Sciences (University of Nottingham).


\begin{thebibliography}{9}
\bibitem{reviewdevoret}
M. H. Devoret and R. J. Schoelkopf, \textit{Science} \textbf{339}, 1169 (2013).
\bibitem{reviewnori}
J. Q. You and F. Nori, \textit{Nature} \textbf{474}, 589 (2012).
\bibitem{reviewwilhelm} J. Clarke and F. K. Wilhelm, \textit{Nature} \textbf{453} 1031 (2008).
\bibitem{annealing} S. Boixo, T. F. R{\o}nnow, S. V. Isakov, Z. Whang, D. Wecker, D. A. Lidar et al. \textit{Nature Phys.} \textbf{10}, 218 (2014). 
\bibitem{cavity}
A. Houck, H. Tureci, J. Koch, \textit{Nature Physics}, \textbf{8} 292-299 (2012).
\bibitem{simulatorfermion} R. Barends, L. Lamata, J. Kelly, L. Garc\'ia-\'Alvarez, A. G. Fowler, A. Megrant, E. Jeffrey et al. \textit{Nature Comm.} \textbf{6}, 7654 (2015).
\bibitem{advances} V. Giovannetti, S. Lloyd and L. Maccone, \textit{Nature Phot.} \textbf{5}, 222 (2011).
\bibitem{ligogw} The LIGO scientific collaboration, \textit{Nature Phys.} \textbf{7}, 962 (2011).
\bibitem{komarnet} P. K\'om\'ar, E. M. Kessler, M. Bishof, L. Jiang, A. S. S{\o}rensen, J. Ye and M. D. Lukin, \textit{Nature Phys.}  \textbf{10}, 582 (2014).
\bibitem{casimirwilson}
C. M. Wilson, G. Johansson, A. Pourkabirian, M. Simoen, J. R. Johansson, T. Duty, F. Nori and P. Delsing, \textit{Nature} \textbf{479}, 376-379 (2011).
\bibitem{casimirtheory} J. R. Johansson, G. Johansson, C.M. Wilson, F. Nori, \textit{Phys. Rev. A} \textbf{82},052509 (2010).
\bibitem{liberatoarray} R. Stassi, S. De Liberato, L. Garziano, B. Spagnolo and S. Savasta, \textit{Phys. Rev. A} \textbf{92}, 013830 (2015).  
\bibitem{borjatunable} B. Peropadre, D. Zueco, F. Wulschner, F. Deppe, A. Marx, R. Gross et al. \textit{Phys. Rev. B} \textbf{90}, 134504 (2013).
\bibitem{wallraffsqueezing}C. Eichler, D. Bozyigit, C. Lang, M. Baur, L. Steffen, J. M. Fink et al. \textit{Phys. Rev. Lett.} \textbf{107}, 113601 (2011).
\bibitem{nonclassicaldce} J. R. Johansson, G. Johansson, C. M. Wilson and F. Nori, \textit{Phys. Rev. A} \textbf{87}, 043804 (2013).
\bibitem{braun} O. Pinel, P. Jian, N. Treps, C. Fabre and D. Braun, \textit{Phys. Rev. A} \textbf{88}, 040102 (R) (2013). 
\bibitem{adesso}
G. Adesso, Phys. Rev. A 90, 022321 (2014).
\bibitem{marianmarian} P. Marian, T. A. Marian, \textit{Phys. Rev. A} \textbf{86}, 022340 (2012).
\bibitem{cavity2} D. L. Underwood, W. E. Shanks, J. Koch, A. Houck, \textit{Phys. Rev. A} \textbf{86}, 023837 (2012).

\end{thebibliography}
\end{document}